# A direct probe of the variability of Coulomb correlation in Fe-pnictide superconductors

P. Vilmercati[1,2], C. Parks Cheney[1], F. Bondino[3], E. Magnano[3], M. Malvestuto[4], M.A. McGuire[5], A.S. Sefat[5], B.C. Sales[5], , D. Mandrus[5,6], D.J. Singh[5], M.D. Johannes[7], N. Mannella[1]*

1. Department of Physics and Astronomy, University of Tennessee Knoxville, 1408 Circle Drive 37996, Knoxville TN
2. Joint Institute of Advanced Materials, University of Tennessee Knoxville
3. Laboratorio TASC, IOM-CNR, S.S. 14 km 163.5, Basovizza, I-34149 Trieste, Italy
4. Sincrotrone Trieste S.C.p.A, Area Science Park, S.S. 14 km 163.5, Basovizza, I-34149 Trieste, Italy
5. Materials Science and Technology Division, Oak Ridge National Laboratory, Oak Ridge, Tennessee 37831-6056, USA
6. Department of Material Science and Engineering, University of Tennessee Knoxville, 1512 Middle Drive Knoxville TN 37996
7. Center for Computational Materials Science, Naval Research Laboratory, Washington, D.C. 20375

*nmannell@utk.edu

**ABSTRACT**

We use core-valence-valence (CVV) Auger spectra to probe the Coulomb repulsion between holes in the valence band of Fe pnictide superconductors. By comparing the two-hole final state spectra to density functional theory calculations of the single particle density of states, we extract a measure of the electron correlations that exist in these systems. Our results show that the Coulomb repulsion is highly screened and can definitively be considered as weak. We also find that there are differences between the 1111 and 122 families and even a small variation as a function of the doping, *x*, in $Ba(Fe_{1-x}Co_x)_2As_2$. We discuss how the values of the hole-hole Coulomb repulsion obtained from our study relate to the onsite Coulomb parameter "U" used in model and first principles calculations based on dynamical mean field theory, and establish an upper bound for its effective value. Our results impose stringent constraints on model based phase diagrams that vary with the quantity "U" or "U/W" by restricting the latter to a rather small range of values.

**INTRODUCTION**:

The degree of correlation in the iron-based superconductors has been widely discussed and debated since the discovery of this class of compounds [1]. Because of the nearby magnetic state and the elimination of electron-phonon coupling as the pairing mechanism [2,3], comparisons to the strongly correlated cuprates were initially widespread. However, the pnictide materials are uniformly metals throughout their doping/pressure phase diagrams and conventional density functional methods, which typically fail for correlated systems, were shown to capture many, though not all, of the electronic properties, such as Fermi surfaces and shapes of the density of states [4,5,6,7,8,9]. Spectroscopic measurements reveal that the Fe spectra have not displayed definite signatures commonly associated with a localized character of the $3d$ electrons and indicative of strong on-site Coulomb repulsion as found for example in the Cu $2p$ core level photoemission spectra of the cuprates high temperature superconductors or Fe oxides. The Fe spectra are characterized by lineshapes more akin to those of Fe metal and intermetallic compounds, and by a high density of states at the Fermi level, in stark contrast to the case of correlated oxides [10,11,12,13,14,15,16]. Angle Resolved Photoemission results suggest the same conclusions: The spectra are renormalized, with a narrowing of the $d$ bands, but Hubbard bands are not found [17,18]. Thus it is becoming accepted that the Fe-based pnictide superconductors are moderately to weakly correlated systems. Nonetheless, it is also clear that electronic interactions are not negligible, and therefore it is useful to have quantitative measurements of the level of correlation that exists. Moreover, the existence of more than one distinct family within the Fe-based superconductors makes a comparison of correlation between families and as a function of doping and/or pressure potentially useful in understanding observed differences in superconducting, magnetic and other behaviours.

Discussion of correlation in Fe-based superconductors often takes place in the context of the "U" – the parameter that controls the onsite Coulomb interaction between localized electrons in the Hubbard model, a version of which is imported into density functional theory as "LDA+U", or in dynamical mean field theory (DMFT) calculations [19]. Attempts to determine its value, both experimentally and theoretically, have produced a wide range of results that have been used to categorize the Fe-based superconductors as everything from weakly to moderately or even strongly correlated systems [16,20,21,22,23,24,25]. Much of this variation comes from ill-defined use of the term "U", which has quantitative meaning only in the context of the specific model being discussed. It is particularly difficult to find strict correspondence between experimental measurements, in which the orbitals/sites involved cannot always be controlled, and theoretical models in which the orbitals/sites are limited and sharply defined, and vary with the model.

Here, we present an experimental determination of the Coulomb repulsion between two holes in the valence band (VB) states derived from Fe orbitals of several Fe-based pnictides superconductors, providing a direct measure of the effective Coulomb correlation strength. Core-valence-valence (CVV) Auger spectra are compared to density functional theory (DFT) calculations to extract a quantitative measurement of local hole-hole repulsion. Recognizing the substantial itinerancy in these systems, and to avoid the impression of directly measuring the Hubbard "U", we refer to our measured value of the screened repulsion between holes in the VB as $\mathcal{U}$. We later discuss whether the measured value can be adapted for use as the onsite "U" parameter in model or DFT calculations. We compare values between the 122 (BaFe$_2$As$_2$) and 1111 (CeFeAsO) families and within the 122 family as a function of doping. Our results indicate a clear upward shift from $\mathcal{U} \approx 1.9$ eV in the Ba-based 122 compounds to $\mathcal{U} \approx 2.5$ eV in the 1111 compound. The $\mathcal{U}$ value also shows light, non-monotonic variation as a function of cobalt doping in $BaFe_{2-x}Co_xAs_2$. This may have significance for the onset of the superconducting state, which is currently believed to be mediated by electronic interactions, most likely in the spin channel (e.g. spin-fluctuations) [26,27]. Clearly, the balance between magnetic order and superconductivity will depend on the details of the electronic structure.

**DESCRIPTION OF THE AUGER PROCESS AND APPROACH TO OBTAIN U:**

It has been established for a long time that the analysis of the core-valence-valence (CVV) Auger spectral lineshapes can give valuable insights into the nature of the VB states and the occurrence and magnitude of electron correlation effects [28,37]. The initial state for the Auger decay consists of the creation of a hole in a deep core level. The system restores a minimum energy configuration by filling the core hole with an electron occupying a higher energy level and promoting a second electron to the continuum, the Auger electron. The Auger effect thus leaves the system in a two-hole final state. When the final state consists of both holes in the VB, the Auger process is labeled as CVV.

In absence of any correlation effects, if $E_B(V_1)$ and $E_B(V_2)$ denote the binding energy (BE) of two electrons in the VB, the kinetic energy (KE) of the Auger electron in a CVV Auger transition is given by $KE = E_B(C) - E_B(V_1) - E_B(V_2)$, where $E_B(C)$ denotes the BE of the core level involved. In a metallic system, the maximum KE of a CVV Auger transition occurs when the two holes are located at the Fermi level (FL), and should thus be equal to the Fermi-referenced BE of the core level, since in this case $E_B(V_1) \approx E_B(V_2) \approx 0$. If $W$ denotes the bandwidth, the minimum KE of the Auger spectrum should be given by $E_B(C) - 2W$, corresponding to the configuration in which the two holes are located at the bottom of the VB. The width of the Auger spectrum amounts thus to $2W$, i.e. twice the bandwidth. In fact, when one considers all the possible pairs of electrons in the VB participating to the Auger decay, it is expected that in absence of correlations the lineshape of the CVV Auger spectrum resembles the self-convolution of the occupied local density of states (SCDOS), with spectra being referred to as band-like [28]. As shown by numerous investigations, this simple picture often breaks down for 3d narrow-band metals [29,30,31,32,33,34,35]. In this case the Auger spectra do not resemble the SCDOS, but are quasi-atomic like, with line shapes describable in terms of atomic models [29,30,31,32,33,34,35]. In a series of independent works, Cini and Sawatzky recognized that these quasi-atomic like Auger spectra are a direct result of the occurrence of strong electron correlation effects [36,37,38].

The Cini-Sawatsky Theory (CST) provides the basis for the current description of CVV Auger spectra in solids. Taking into account effects beyond the single particle approximations, the measured KE of CVV Auger electrons can be written

(1) $\qquad KE(C,V_1,V_2;X) = E_B(C) - E_B(V_1) - E_B(V_2) - \mathcal{U}(V_1,V_2;X)$

where $\mathcal{U}(V_1,V_2;X)$ is the *effective* Coulomb interaction or correlation energy of the two holes in the final state X. $\mathcal{U}$ can be written as $\mathcal{U} = \mathcal{U}^0(V_1,V_2;X) - R(V_1,V_2;X)$, an expression indicating that the bare intra-atomic Coulomb energy $\mathcal{U}^0(V_1,V_2;X)$ of interaction between the two holes in the final state is reduced in a solid by a variety of relaxation and screening effects described by $R(V_1,V_2;X)$, which will be discussed in more details below [34]. The CST provides the distribution D(E) of the Auger electrons when correlations are at work in terms of the Auger distribution $D^0$(E) obtained in the absence of correlations, thus providing a quantitative method to evaluate $\mathcal{U}$:

$$(2) \quad D(E) = \frac{D^0(E)}{[1-UF(E)]^2 + U^2 D^0(E)^2} ,$$

where $D^0(E) = \int_{VB} n(E-\Delta)n(E+\Delta)d\Delta$ is the SCDOS, i.e. the self-convolution of the occupied local DOS $n(E)$, and $F(E) = P \int_{-\infty}^{\infty} \frac{D^0(\Delta)d\Delta}{E-\Delta}$ is the Hilbert transform of $D^0(E)$. When correlation effects are not present, i.e. $\mathcal{U} = 0$, Eqn. (2) shows that $D(E)$ coincides with $D^0(E)$, as in the most simple case [39].

As described in equation (1), because of conservation of energy in the Auger process, for each couple of valence electrons $V_1$ and $V_2$ involved in the Auger decay the effect of the hole-hole interaction $\mathcal{U}$ in the final state is to decrease the KE of the outgoing Auger electron and redistribute the spectral weight of the Auger lineshape. A comparison between the calculated SCDOS and the measured Auger spectrum plotted on the so called two-hole scale, obtained by subtracting the BE of the initial core hole from the measured kinetic energy of the Auger spectrum, allows an estimate of the value of $\mathcal{U}$ averaged over all of the possible hole-hole pairs and final states X. Specifically, the average value of $\mathcal{U}$ is given by the difference between the average energies, hereafter referred to as centroids, of the Auger spectrum and the SCDOS $D^0(E)$. The latter are identified with the first moment of the Auger and SCDOS spectral distributions according to $\langle E \rangle = \frac{\sum_i I(\varepsilon_i)\varepsilon_i}{\sum_i I(\varepsilon_i)}$, where $I(\varepsilon_i)$ denotes the intensity of the spectrum evaluated at the energy value $\varepsilon_i$ on the two-hole scale.

Another independent determination of $\mathcal{U}$ is provided by fitting the spectra with the $D(E)$ profile calculated with Eq. 2 treating $\mathcal{U}$ as a best fit parameter. For $\mathcal{U}/W < 1$, the Auger spectra are band-like, i,e. they resemble the SCDOS profile. As the ratio $\mathcal{U}/W$ increases from zero, the profile $D(E)$ deviates from the SCDOS with a pronounced skewing towards lower KE. In the limit $\mathcal{U}/W = 1$, a bound state appears, completely separated from the VB states. For $\mathcal{U}/W > 1$, more spectral weight is transferred to the bound state at the expense of the band-like portion of the spectrum, eventually reaching the quasi-atomic limit in which the spectral lineshapes are similar to those found in isolated atoms, with no band-like contributions [39]. This limit corresponds to a vanishing denominator in Eqn. 2 and represents a two-hole resonance in which the two holes form a localized state via their strong repulsion.

The initial formulation of the CST is rigorously valid only for full electron bands, and it neglects core-hole lifetime effects. Further extensions of the CST have allowed the description of CVV Auger spectra in more complex cases that include the occurrence of partially filled band, the off-site contribution to electron correlations, and dynamical screening of the core hole [40,41,42,43,44,45,46,47]. In the band-like limit ($\mathcal{U}/W < 1$) the contribution from off-site correlations or complications arising from partially filled band with low hole-number are less important than in the atomic-limit ($\mathcal{U}/W > 1$) [41,42,43,44,47]. A proper description of the theory for the whole *3d* series requires taking into account the dynamical screening of the core hole by the valence electrons in the initial state [40,45,46,47]. Because of the hypothesis of filled bands, the initial formulation of the CST (cf. Eqn. 2) does not take into account the spectrum of the excited states of the system. In the description of the Auger process, the initial state is the un-relaxed final state of the photoemission process, i.e. following the creation of the core hole. This means that all the possible photoemission final states, including shake-up and shake-off satellites, have to be considered. Because of the occurrence of these final states, the Auger spectrum is followed by a long tail in the low kinetic energy side (intrinsic process). In addition, the Auger electrons can lose energy by shake up/shake off processes, producing a long tail in the Auger spectrum on the low kinetic energy side (extrinsic process). The electrons emitted according to these two processes are indistinguishable in ordinary measurements, but their contribution can be disentangled with the use of time coincidence techniques [48]. When comparing the SCDOS with the experimental Auger spectrum, a suitable way to take these effects into account is by convolving the lineshape *D(E)* obtained in Eqn. 2 with the measured core level photoemission spectrum [45,46,47].

## EXPERIMENTAL AND METHODS

Polycrystalline samples with composition $CeFeAsO_{0.89}F_{0.11}$ (optimally doped, CFAO-OD) were grown with a standard solid-state synthesis [49,50]. Magnetic susceptibility measurements show the onset of superconductivity near 38 K. High quality $Ba(Fe_{1-x}Co_x)_2As_2$ and $SrFe_2As_2$ single crystals were grown out of FeAs flux using the procedure described elsewhere [51]. For the undoped materials $BaFe_2As_2$ (BFA) and $SrFe_2As_2$ (SFA) the antiferromagnetic to paramagnetic transition temperatures $T_N$, determined by resistivity and magnetization measurements, occur at 140K and 190K, respectively. The fraction of Co content *x* in the formula unit $Ba(Fe_{1-x}Co_x)_2As_2$ was determined with electron probe microanalysis carried out with a JEOL JSM-840 scanning electron microprobe. These materials will be described in the following by their actual compositions with doping levels *x* = 6%, 8%, 12%, 22%, corresponding to under-doped (UD), optimally doped (OD), over-doped (OvD) and heavily over-doped (HD), respectively.

The Auger and photoelectron spectra were recorded on the BACH beamline at the Elettra Synchrotron Facility [52]. Several samples from different batches have been measured at 100K in a pressure better than $4 \times 10^{-10}$ mbar after being fractured (CFAO-OD) or cleaved ($Ba(Fe_{1-x}Co_x)_2As_2$ and SFA) *in situ*. The experimental resolution amounts to $\approx$ 350 meV.

DFT calculations were performed within the local density approximation using the general potential linearized augmented planewave (LAPW) method as implemented in the Wien2k code, without including magnetism [53,11]. The projections of the electronic densities of states were obtained by integration within the LAPW spheres. For Ba(Fe$_{1-x}$Co$_x$)$_2$As$_2$, the As position was input based on the experimental crystal structure of Q. Huang et al. at 175 K for pure BaFe$_2$As$_2$ [54]. The Co doping was modeled using the virtual crystal approximation (VCA). Calculations for CFAO-OD and additional details are reported in ref. [12]. The validity of the VCA was carefully checked by performing a reference calculation in a supercell that included real Co atoms at a 25% doping level and a full relaxation of the structure with their inclusion. Differences between the calculated centroid in VCA with identical doping and in the supercell were less than 150 meV. This maximum discrepancy has been included in the error bars accompanying our data.

## RESULTS & DISCUSSION

Fig. 1 shows the Fe2p$_{3/2}$-valence-valence Auger Fe(2p$_{3/2}$VV) spectra for several FeSC compounds. The maximum KE of the Auger spectra and the BE of the Fe2p$_{3/2}$ core levels have very similar values. Since the maximum KE of a CVV Auger transition is equal to the Fermi-referenced BE of the core level corrected by $\mathcal{U}$ (cf. Eqn. 1), these data immediately suggest that the magnitude of $\mathcal{U}$ in these materials is not large. This is consistent with the fact that the full width at half maximum (FWHM) of the Auger spectrum is $\approx$ 6 eV, which immediately indicates that the spectrum is band-like, thus further confirming that the magnitude of the Coulomb interaction $\mathcal{U}$ in these materials is modest. The magnitude of the shifts and the shape of the Auger lines samples are markedly different as compared to those measured in strongly correlated systems with $\mathcal{U}/W \simeq 1$ such as cobaltates, cuprates and manganites [55,56,57,58].

A first determination of the values of $\mathcal{U}$ consists in the evaluation of the energy difference between the centroids of the SCDOS $D^0(E)$ and the Auger spectra plotted on the two-hole scale. Since it represents the density of two hole states in absence of hole-hole interactions, D$^0$(E) cannot be determined from photoemission, but should be calculated as the self convolution of the calculated one-particle density of states (DOS) [37]. Moreover, in the case of compounds, the measured VB derives from valence orbitals belonging to the different elements (Sr, Ba, Fe and As in the 122 compounds, and Ce Fe As and O in the 1111 compounds). As the Fe(2p$_{3/2}$VV) Auger transition involves the Fe 2p$_{3/2}$ core hole, we consider the calculated Fe 3d partial DOS (Fe$_{3d}$ p-DOS), as shown as an example for BFA and CFAO-OD in Fig. 2a and Fig. 2b, respectively. The demonstrated possibility of well matching the occupied DOS determined from standard DFT calculations to valence bands photoemission spectra [16,12] gives us confidence in this procedure. To account for the lifetime of the VB states and the experimental resolution, the Fe$_{3d}$ p-DOS has been convolved with a mixed Lorentzian and Gaussian broadening, resulting in a profile denoted as Fe*$_{3d}$. As shown in the bottom part of Fig. 2, a direct comparison between the experimental VB and the obtained Fe*$_{3d}$ profile outlines the optimal match between Fe*$_{3d}$ and the data points in proximity of the top of the VB. We identify the $\approx$ 6 eV range of the calculated Fe*3d profile as the bandwidth $W$ relevant to the following discussions. The parameter $\mathcal{U}$ has been estimated self-consistently by directly comparing the experimental data with D$^0$(E), and with the lineshape D(E) calculated with Eqn. 2, with results shown in Fig. 3. The SCDOS $D^0(E)$ has been evaluated by the self-convolution of the Fe*3d profile.

The comparison between the experimental data and $D^0(E)$ allows a determination of $\mathcal{U}$ by evaluating the energy difference between the centroids of $D^0(E)$ and the Auger spectra plotted on the two-hole scale (Fig. 3a). The magnitude of $\mathcal{U}$, identified as the difference between the centroids, is ≈ 2eV, with values ranging from a minimum of 1.66 eV (BFA-OvD) to a maximum value of 2.53 eV (CFAO-OD). With a $Fe_{3d}$ states-derived bandwidth W ≈ 6 eV, the $\mathcal{U}/W$ ratio ≈ 0.3 identifies the Auger spectra as band-like ($\mathcal{U}/W < 1$), thus placing the pnictides in the category of weakly or at best moderately correlated materials, in agreement with the estimates provided in Ref. [16]. This is also in excellent agreement with the determination of the physical range of parameters in Ref. [59] obtained by comparing experimental and theoretical results.

We obtain consistent results when $\mathcal{U}$ is determined with the CST, provided that the core hole screening effects are taken into account. This is not unexpected, since it has been shown that in the band-like limit ($\mathcal{U}/W < 1$) the contribution from off-site correlations or complications arising from partially filled bands with low hole-number are less important than in the atomic-limit ($\mathcal{U}/W > 1$) [41,42,43,44,47], and that core hole screening effects are crucial for describing open shell systems [40,45,46,47]. To account for the core hole broadening in the Auger $Fe2p_{3/2}VV$ transition and final state effects upon creation of the core hole such as shake up/shake off processes, the expression for the lineshape of the Auger spectrum D(E) given in Eqn. 2 has been convolved with the Fe $2p_{3/2}$ photoemission spectra, whose lineshape consists of a main peak with an asymmetric tail at higher BE (cf. Fig. 1). Because of the pronounced asymmetry of the Fe $2p_{3/2}$ spectra, the convolution operation shifts the maximum of the peak of the resulting profile to higher values as compared to the D(E) profile alone. The fits of the Auger spectra with the resulting profiles are excellent, especially taking into account that the background on the high KE side (i.e. lower values on the two-hole scale) of the $Fe(2p_{3/2}VV)$ Auger spectra are enhanced by the tails of an adjacent peak corresponding to a $Fe(2p_{3/2}VV)$ Koster-Kronig proceeded Auger decay [60]. The values of the parameter $\mathcal{U}$ (Fig. 3b) that optimize the fits are found to be quite similar (i.e. within 10%) to those determined independently by evaluating the difference of the centroids of the spectra and the $D_0(E)$ profile (cf. Fig. 3a), with the only exception being BFA-OD, where agreement is within 20%.

Fig. 4 shows the $\mathcal{U}$ values obtained from comparing the experimental spectra with $D_0(E)$ and $D(E)$. The more detailed doping dependence for the doped BFA system shows that there are small, but significant, non monotonic variations in the magnitude of $\mathcal{U}$, with maximum and minimum values for the optimally doped and over-doped composition, respectively. It is impossible, on the basis of our experiment, to disentangle the contributions of the terms $\mathcal{U}^0(V_1,V_2;X)$ and $R(V_1,V_2;C)$ from the non monotonic variations in the magnitude of $\mathcal{U}$. Although we cannot exclude a priori that the bare Coulomb repulsion $\mathcal{U}^0(V_1,V_2;X)$ between the two holes located in the VB, this quantity is intra-atomic, and thus we do not expect this to change significantly with doping. Rather, the doping dependence of the values of $\mathcal{U}$ indicates the occurrence of different relaxation and screening effects, as described by R(V$_1$,V$_2$;C), which depends sensitively on doping. The total relaxation energy $R$(V$_1$,V$_2$;X) describes the additional relaxation energies of the two-hole final state above and beyond the one-hole dynamic relaxations already contained in the expression of the BE's (cf. Eq. 1). The one-hole dynamic relaxations correspond to the amount of energy by which the BE's are lowered from the frozen orbital energies estimated according to Koopman's theorem through relaxation of the passive electrons during electrons emission [61]. These relaxations are referred to as "dynamic" to emphasize that they occur concomitantly to processes in which electrons are emitted. The total relaxation energy $R$(V$_1$,V$_2$;X) consists in the sum of two contributions, $R$(V$_1$,V$_2$;X) = R$_S$(V$_1$,V$_2$;X) + ΔR(V$_1$,V$_2$). The term ΔR(V$_1$,V$_2$) = R$^{V1}$(V$_2$) – R(V$_2$) represents the difference between the dynamic relaxations upon emission of one electron from level V$_2$ in the presence (R$^{V1}$(V$_2$)) or in absence (R(V$_2$)) of a hole in level V$_1$ [62]. The term R$_S$(V$_1$,V$_2$) denotes the static relaxation energy, i.e. the amount of energy by which the frozen BE (i.e. Koopman's value) of the V$_2$ electron is reduced because of the relaxation of the passive electrons induced by the hole in level V$_1$. The term "static" emphasizes that this is an initial state relaxation, distinguished from the dynamic relaxations occurring in conjunction with emission of the V$_1$ and V$_2$ electrons [61]. The total relaxation energy $R$(V$_1$,V$_2$;X) is thus a complex term, which does not offer a simple description in terms of quantities that are directly measurable in experiments.

These results show conclusively that the interaction between holes in the VB of these pnictide compounds is highly screened and that these systems can be considered as weakly correlated from the point of view of the strength of the on-site Coulomb repulsion. It is worth noting that our calculations, if anything, err on the high side of the $\mathcal{U}$ value. We have selected out only the *Fe d* states in a radius around the Fe ion in real space for inclusion in our calculation of *D$_0$(E)*. Given that the *Fe d* and *As p* states are mixed in this itinerant system, an alternate approach would be to include both of them, provided they are within the same real space radius in an attempt to approximate the hybrid orbitals that compose the VB in the crystal. Since the *As p* states are lower in energy than the *Fe d*, this would invariably shift the centroid of the calculated *D$_0$(E)* towards the low KE side (i.e. higher values on the two-hole scale), and closer to the centroid of the experimental spectra, thereby reducing the value of $\mathcal{U}$. Thus, our values are an upper bound on the correlation, further emphasizing its weak character.

Now we discuss the relationship between the screened Coulomb repulsion that we measure, $\mathcal{U}$, and the Hubbard "U". As stated earlier, "U", is a theoretical construct that is model-dependent. Therefore, the relevant question becomes: which model (or basis set), if any, best corresponds to the orbitals probed by our experiment? The problem with giving a direct and clear answer to this question is that the centroid of the Auger spectrum is a real, observable quantity, whereas the calculated centroid of the SCDOS depends on how the local density of states is calculated. An ideal calculation (and corresponding model) would take into account precisely the set of orbitals occupied by the two holes involved in the final CVV state, but a full knowledge of how to describe these orbitals is unavailable. In his original formulation Sawatzky reasonably assumed Wannier functions localized to the relevant atom [37]. It is nicely illustrated in Ref. [15, 23] that this method is non-unique for a multi-band system, as the "U" value corresponding to a Wannier function-based model depends strongly on how many Wannier orbitals are included. In this work, we used a projection method to select out contributions from states within a sphere around the Fe atom that have *d*-like symmetry (see methods section). The resulting $\mathcal{U}$ should therefore be conceived of as the difference between the SCDOS resulting from *specifically* these selected orbitals and the measured Auger spectra. If instead we had included in our calculations states that are hybridized with *As p* character but still within a sphere around the Fe atom, the centroid of the SCDOS would shift toward the low KE side (i.e. higher values on the two-hole scale), and closer to the centroid of the experimental spectra, thereby reducing the value of $\mathcal{U}$. Calculating the local density of states via a Wannier function approach would yield yet again different values. Thus, our $\mathcal{U}$ should not be understood to provide a quantitative value for any specific Hubbard or Hubbard-based model requiring a "U". However, since we have used a highly restrictive method of choosing the orbitals that comprise our SCDOS, and since including any other states would shift the centroid of the calculated SCDOS to higher values on the two-hole scale, our value of $\mathcal{U}$ provides an upper bound for the Hubbard "U". The main message of this work, then, is that the Coulomb correlation as defined by an on-site Coulomb repulsion beyond the DFT level is weak or at best moderate (and certainly not strong). As a secondary point, we note that its strength varies with the different compounds within families and perhaps very lightly as a function of doping. This constrains the theoretical description of phase diagrams that vary with the quantity "U" or "U/W" by restricting U" or "U/W" to a rather small range of values. On the other hand, this does not address other types of electron correlation, specifically spin-fluctuations that couple to the *d*-electrons and can lead to renormalizations and shifts in spectral weight within the d-bands, with coupling through the exchange correlation (Hunds rule) term as in Refs. [27,63,64].

## CONCLUSIONS

We have obtained the CVV Auger spectra of CeFeAs$_{O0.89}$F$_{0.11}$, SrFe$_2$As$_2$, and Ba(Fe$_{1-x}$Co$_x$)$_2$As$_2$ for a range of different doping levels *x*. We compare the two-hole final state Auger spectra with DFT calculations of the convolution of the single particle density of states. We extract a measure of the screened Coulomb repulsion between two holes in the valence band. Our results show that this screening is considerable and that all the systems we have investigated can be definitively considered as having weak on-site Coulomb repulsion. We also find that there are differences between the systems, most prominently between the 1111 and 122 families, but also weakly within the 122 family as a function of doping. We argue that, although the quantitative values we report are not suitable for direct use in model calculations of the Hubbard type, they provide constraints on the type of model that can be regarded as describing these materials.

## ACKNOWLEDGEMENTS

Work at UTK and Elettra Sincrotrone Trieste was supported by NSF grant DMR-1151687. Work at ORNL was supported by the Department of Energy, BES, Materials Sciences and Engineering Division.

FIGURES and FIGURES CAPTIONS

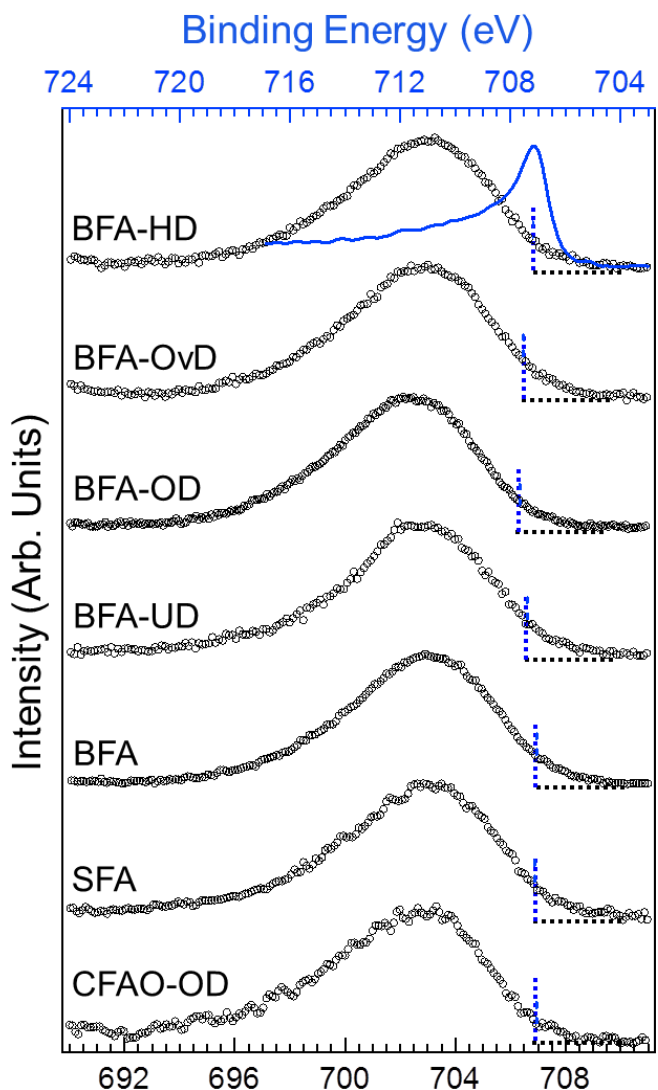

Fig. 1.(Color Online). Fe $2p_{3/2}$VV Auger spectra of different compounds plotted on a kinetic energy scale (bottom scale). The background on the high KE side of the spectra are enhanced by the tails of the adjacent Fe($2p_{1/2}$VV) Auger spectra. The continuous line is the Fe $2p_{3/2}$ core level photoemission spectrum, plotted on a binding energy scale (top scale). A Shirley background has been removed from all of the spectra. The vertical dotted lines indicate the position of the maximum of the Fe $2p_{3/2}$ spectra as determined fits of the spectra with a Doniach-Sunjic profile [65]. These values are referenced to the kinetic energy scale (bottom scale). The proximity of the maximum of the Auger spectra and the BE of the Fe $2p_{3/2}$ core level is indicative of low values of $U$.

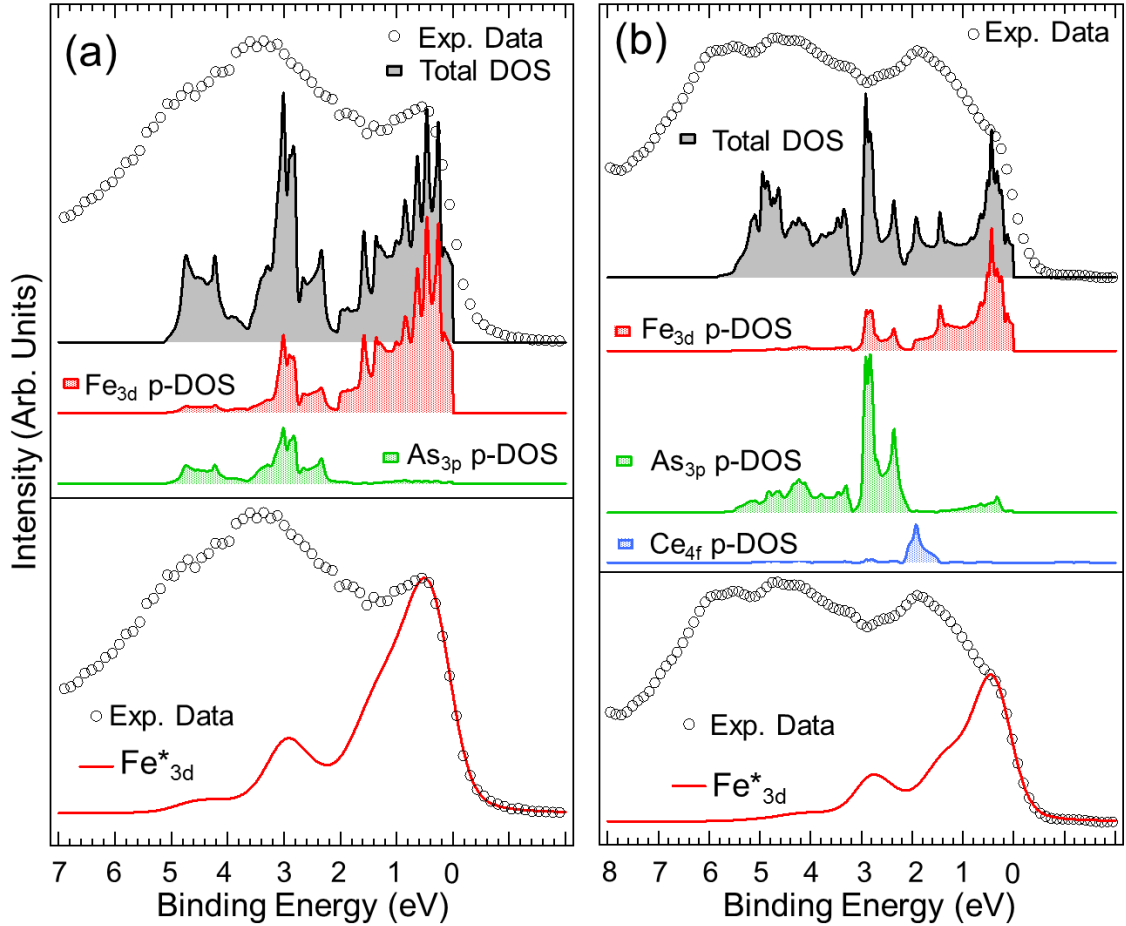

Fig. 2. (Color Online). Valence band photoemission data collected in (a) BFA and (b) CFAO-OD with photon energy of 1065 eV (open circles) compared with the Fe 3d, As 3p and Ce 4f partial density of states (p-DOS) calculated with Density Functional Theory. The lower panels show the valence band data compared to the Fe*$_{3d}$ profile obtained from convolving the Fe 3d p-DOS with a mixed Lorentzian and Gaussian broadening to account for the lifetime of the valence band states and the experimental resolution.

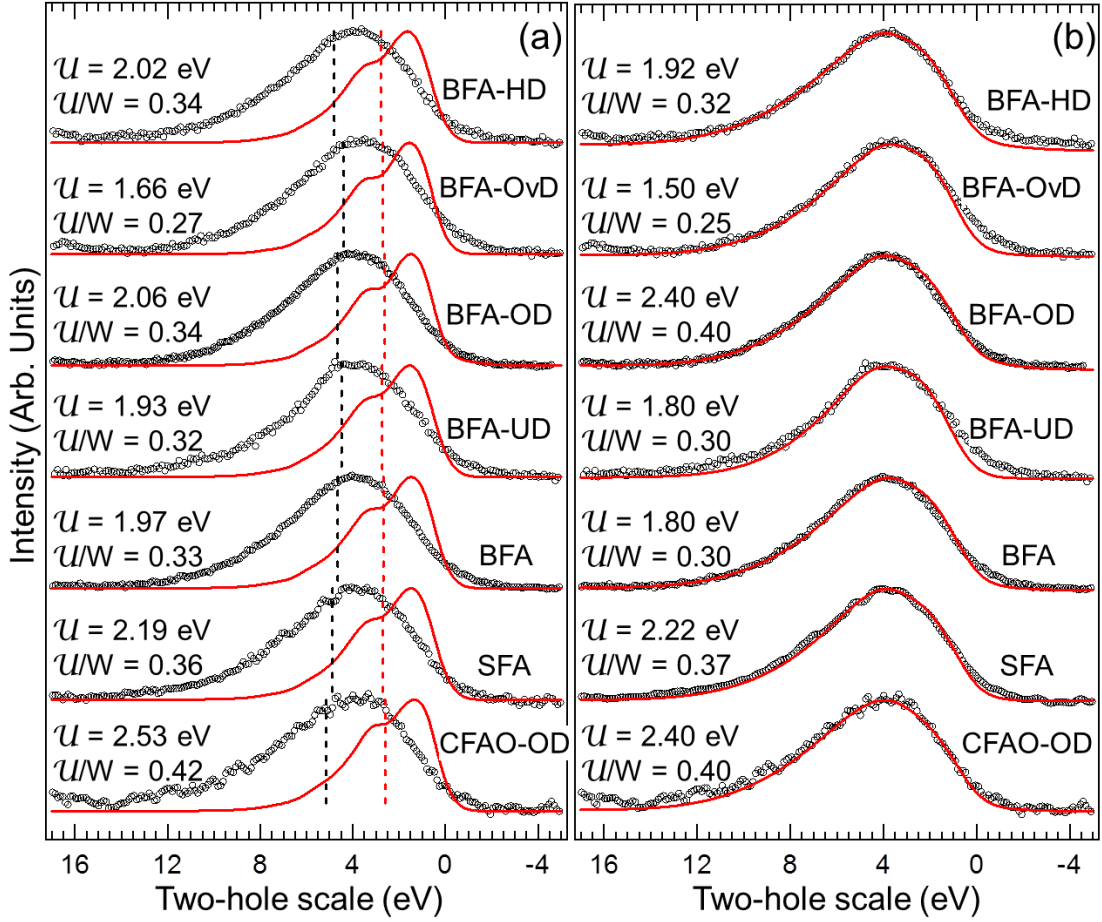

Fig. 3. (Color Online). Determination of the effective energy $U$. (a) Fe2p$_{3/2}$VV Auger spectra and calculated $D^0(E)$ plotted on the two-hole scale for different compounds. The dotted lines denote the centroids (i.e. weighted averages) of the Auger spectra and the $D^0(E)$ lineshape. The energy difference between the centroids provides an experimental assessment of the value of the effective $U$. (b) Fe2p$_{3/2}$VV Auger spectra plotted on the two-hole scale for different compounds compared to the calculated spectra obtained by convolving the $D(E)$ profile (CST, Eqn. 2) with the Fe 2p$_{3/2}$ photoemission spectra to account for the core hole broadening in the Auger Fe2p$_{3/2}$VV transition.

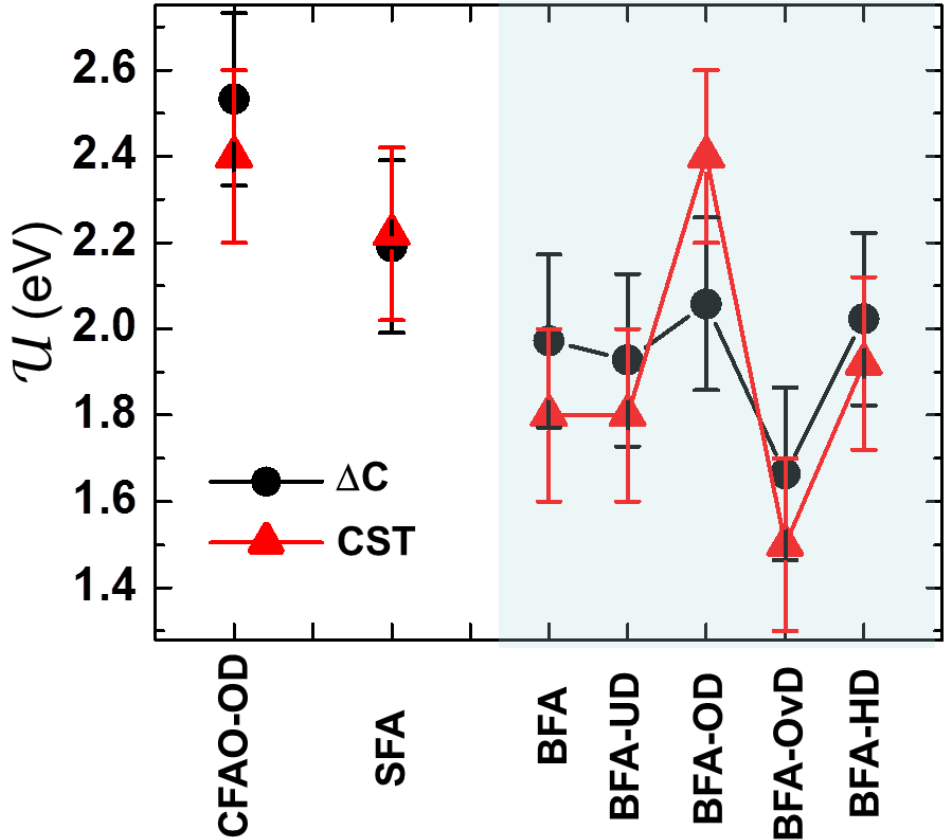

Fig. 4. (Color Online). Tabulation of the effective two-hole repulsion energy U for different compounds. Values are obtained independently by means of evaluating the difference of the centroids of the Auger spectra and the $D^0(E)$ profile (ΔC, circles, black), and by applying the CST as described in the text (SCT, triangles, red). The continuous lines through the data points are a guide to the eye.